\documentclass[natbib]{aastex}

\newcommand{\bvf}{Brunt-V\"ais\"al\"a }
\newcommand{\kic}{KIC 8626021 }
\newcommand{\kicnsp}{KIC 8626021}

\shorttitle{Seven-Period fit of the Kepler DBV}
\shortauthors{Bischoff-Kim}
\usepackage{epstopdf}
\usepackage{array}

\begin{document}

\title{Seven-Period Asteroseismic Fit of the Kepler DBV}

\author{Agn\`{e}s Bischoff-Kim}  
\affil{Penn State Worthington Scranton, Dunmore, PA 18512, USA}
\email{axk55@psu.edu}

\author{Roy H. \O stensen}
\affil{Institute of Astronomy, KU Leuven, Celestijnenlaan 200D, 3001 Heverlee, Belgium}
\email{roy@ster.kuleuven.be}

\author{J.~J. Hermes}
\affil{Department of Physics, The University of Warwick, Coventry CV4 7AL, UK}
\email{j.j.hermes@warwick.ac.uk}

\and

\author{Judith L. Provencal }
\affil{Department of Physics and Astronomy, University of Delaware, Newark DE 19716, USA}
\email{jlp@udel.edu}

\begin{abstract}

We present a new, better-constrained asteroseismic analysis of the helium-atmosphere (DB) white dwarf discovered in the field of view of the original \emph{Kepler} mission. Observations obtained over the course of two years yield at least seven independent modes, two more than were found in the discovery paper for the object. With several triplets and doublets, we are able to fix the $\ell$ and $\rm m$ identification of several modes before performing the fitting, greatly reducing the number of assumptions we must make about mode identification. We find a very thin helium layer for this relatively hot DB, which adds evidence to the hypothesis that helium diffuses outward during DB cooling. At least a few of the modes appear to be stable on evolutionary timescales and could allow us to obtain a measurement of the rate of cooling with monitoring of the star over the course of the next few years with ground-based follow-up.

\end{abstract}


\section{Astrophysical Context}

White dwarfs are the end product of the evolution of around 98\% of the stars in our Galaxy. Buried in their interiors are the records of physical processes that take place during earlier stages in the life of the star. Nuclear reaction rates during the core helium burning phase set the core composition of white dwarfs, while the relative time spent burning hydrogen and helium during the asymptotic-giant-branch (AGB) phase and mass-loss episodes determine the thickness of the helium layer \citep{Lawlor06,Althaus05}. Helium atmosphere white dwarfs (DBs) comprise roughly 20\% of the population of field white dwarfs, with most of the remaining 80\% consisting of their hydrogen atmosphere (DA) cousins. 

The bifurcation into two spectral classes is thought to occur during post-AGB evolution when, in some cases, a very late thermal pulse burns off the residual hydrogen in the envelope, producing a nearly pure helium atmosphere (Iben et al. 1983). Such objects
are then supposed to return to the white dwarf cooling track as PG 1159 stars, which are widely believed to be the precursors of most DB white dwarfs. The hot PG 1159 stars, having recently emerged from the born-again phase, contain envelopes with a nearly uniform mixture of helium (He), carbon (C), and oxygen (O) out to the photosphere \citep{Dreizler98,Herwig99}. As they cool, the helium diffuses upward and gradually accumulates to form a chemically pure surface layer. 

A key prediction of the diffusion models is that, for a given stellar mass, the pure He surface layer will steadily grow thicker as the DB star cools. The only available observational tests of this prediction come from asteroseismology: the study of the internal structure of stars through the interpretation of their pulsation periods. DBs are found to pulsate at effective temperatures ranging between 21,000 K and 28,000 K (Beauchamp et al. 1999; Castanheira et al. 2005), or perhaps hotter, as we find in the present study. At the blue edge of the DBV instability strip, the helium is still in the process of floating up to the surface so that instead of a pure helium layer surrounding the carbon and oxygen core, one has a region where the carbon and helium are still mixed. This leads to a double- layered structure, with the pure He surface layer overlying the remainder of the uniform He/C/O envelope, all above the degenerate C/O core.

GALEXJ192904.6+444708 (a.k.a WD\ J1929+4447 or KIC 8626021 in the \emph{Kepler} input catalog) was found to be a good DBV candidate in a small auxiliary survey undertaken by \citet{Oestensen11}. One month of short cadence (58.8\ s exposures) data was collected by the \emph{Kepler} spacecraft 2010 October and November, revealing a pulsation spectrum with five roughly equally spaced pulsation modes, three of which appeared as evenly split triplets. The Fourier transform and pulsational properties were presented in the discovery paper \citep{Oestensen11} and a seismic study soon afterwards \citep{Bischoff-Kim11b}. One interesting result of the latter was the fact that the object was significantly hotter than previously determined from preliminary spectroscopy. This result was confirmed by an independent study by \citet{Corsico12b}. 

However, the initial asteroseismic studies were limited by the fact that we only had identified five pulsation periods, leading to under-constrained models. Nine further months of observations -- during {\em Kepler} quarters Q10,Q11,Q12 -- revealed two additional periods \citep{Oestensen13}. In this work we present an analysis of the entire {\em Kepler} observations, using the nearly continuous, 2-year data set from Q10 to Q17. We confirm the seven previously reported modes and find additional multiplet structure. With a total of seven modes, we are able to fully constrain the interior of \kic with the models and methods we use. This allows us to add a third data point in the asteroseismic study of elemental diffusion in white dwarfs.

\kic is also of particular scientific significance because it is the second hot DBV with stable periods discovered. Stable hot DBVs are good candidates to study the emission of plasmon neutrinos \citep{Winget04}. An observed rate of change of the periods of any phase-stable mode, combined with models, would allow us to place constraints on the emission of plasmon neutrinos. This concept was attempted with EC\ 20058-5234 \citep{Dalessio10,Bischoff-Kim08c}, but without much success because none of the modes were stable over evolutionary timescales. With the exception of \kic, none of the other known DBVs observed so far show modes stable on evolutionary time scales. On the other hand, evolutionary rates of period changes have successfully been measured for other pulsating white dwarfs, the most notable example being G117-B15A \citep[e.g.][]{Kepler05a}.

In Section \ref{observations}, we summarize the previously published periods for the object and present the new modes discovered when we include data up to Q17. We also comment on the stability of the modes. Section \ref{analysis} is devoted to the asteroseismic analysis of the period spectrum. We extract some mode identifications from the multiplets and then move to grid searches for the best fits. We discuss the implications of our results in Section \ref{discussion}.

\section{Pulsational properties}
\label{observations}

Analysis of one month of data (Q7.2) yielded five periods \citep{Oestensen11}. In a subsequent analysis based on nine months of data taken over Q10, 11, and 12, \citet{Oestensen13} found two more modes. One of these modes, initially named ``$f_{6,0}$'' is not an independent mode, but a combination of $f_{1,-}$ and $f_5$. We list it in table \ref{t1} for completeness, but this mode was not used in the asteroseismic fits. In table \ref{t1}, we also list additional frequencies found in the present analysis.

\begin{figure}
  \includegraphics[scale=0.70,viewport=50 0 50 350]{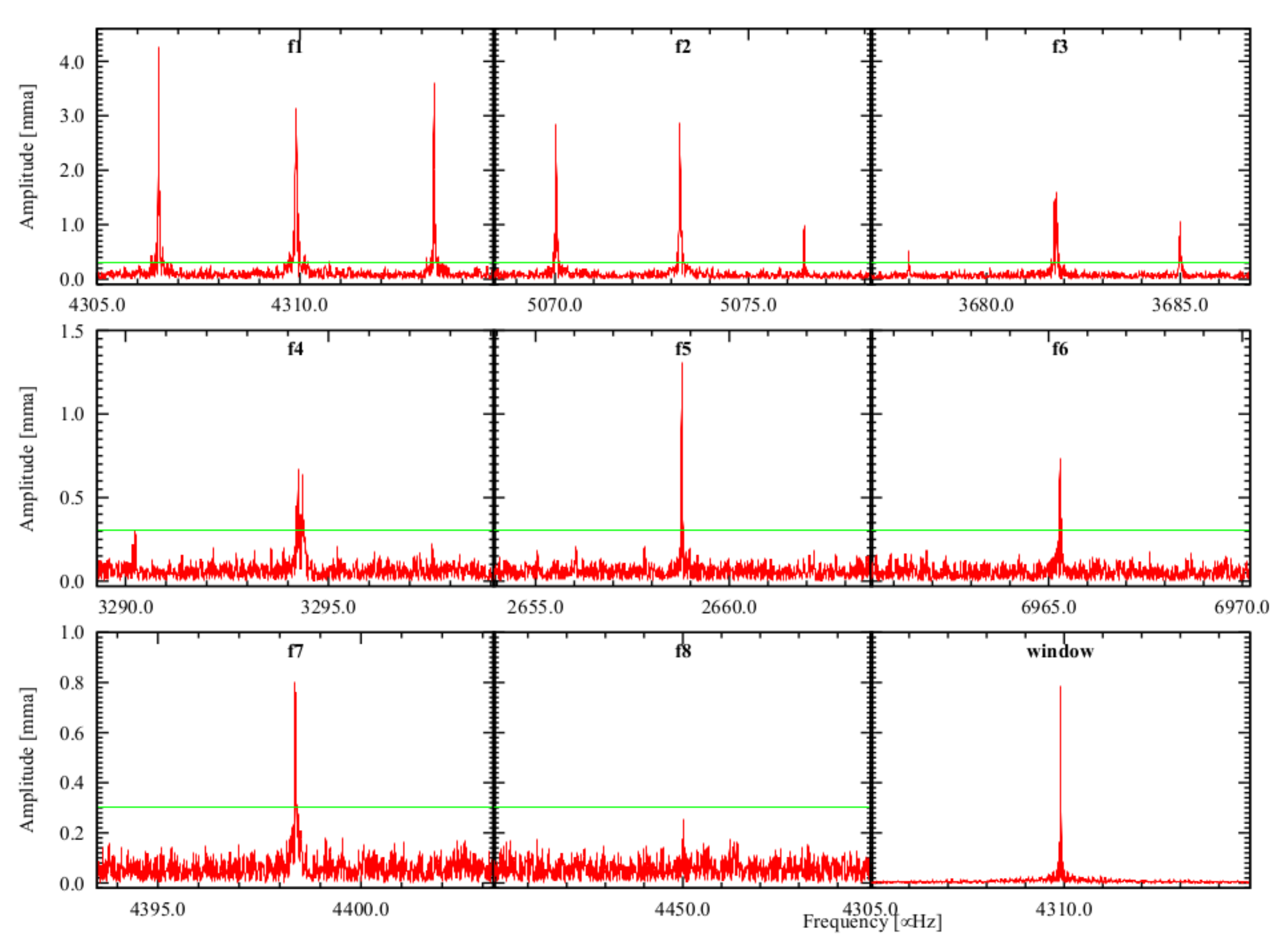}
  \caption{
  \label{f1}
  Fourier peaks of the eight highest-amplitude modes. The triplet structure of $f_1$, $f_2$, and $f_3$ is evident. The green horizontal line denotes the detection threshold set at the 5-$\sigma$ limit of 0.3 mma. $f_8$ is one of the modes we excluded from our initial analysis.
  }
\end{figure}

The exact values of the frequencies listed in Table \ref{t1} are based on the analysis of data collected from Q10 to Q17. The observed periods are simply the inverse of the frequencies. Figure \ref{f1} shows the full Fourier spectrum of \kic based on the nearly continuous data set from Q10 to Q17. The \emph{Kepler} light curve spans 684 days with gaps totaling only about 55 days. The window function is therefore very sharp, and has a resolution 
of ~0.017 $\rm \mu Hz$. With the Fourier transform up to the Nyquist limit of 8496 $\mu$Hz containing 500{,}000 points, one can expect a large number of spurious peaks above a traditional 4-$\sigma$ limit (more than 30 from a Gaussian distribution). We therefore invoke a more strict 5-$\sigma$ limit. The mean level in the FT is 0.06 mma, so we set the detection threshold to 0.3 mma, as indicated by the green horizontal line in Figure1.

In Table \ref{t1}, we list some frequencies that are below our 5-$\sigma$ detection limit, but above the 4-$\sigma$ detection limit. While we are not using them in our initial asteroseismic fits, they are worth listing and discussing how they fit into our analysis. A new peak appears in Q17 data ($f_11$). With the three additional quarters of data, $f_4$ now exhibits a doublet structure with a splitting consistent with that of an $\ell=1$ mode. We use this fact and the previously observed triplet structures of $f_1$, $f_2$, and $f_3$ to guide our asteroseismic fits.

\begin{table}
\begin{center}
\caption{
\label{t1}
A summary of the history of the observed periods and the periods of the best fit models in this work. A blank value in the first two columns means that the mode was not yet apparent in the data. The observed frequencies and corresponding periods listed in the table are those that result from the analysis of the full data set (up to and including Q17). Model 1 refers to asteroseismic fits made using only the periods above the 5-$\sigma$ detection limit (the more conservative model), while Model 2 also includes modes above the 4-$\sigma$ detection limit.
}
\renewcommand{\arraystretch}{0.8}
\begin{tabular}{ccclccccllr}
\tableline\tableline
\multicolumn{3}{c}{Mode Label} 							& Frequency 	& Amplitude				& \multicolumn{3}{c}{Periods (s)} 	& 
\multicolumn{3}{c}{Mode ID} \\
Q7.2\tablenotemark{1} & Q12\tablenotemark{2} & Q17\tablenotemark{3}	& ($\mu$Hz) & (mma) & Observed &  Model 1 &	Model 2 &	 $\ell$  & $k$  & m\\
\tableline
		&			& $f_{11,0}$	& 6981.3 		& 0.38	& 143.24 	& 142.68	& 142.68	&2			 	& 4 		& 0		\\
		& $f_{6,0}$	& $          $ 	& 6965.29  	& 0.74	& 143.57 	& \multicolumn{5}{c}{combination mode ($f_1+f_5$)}	\\
$f_{2,+}$ 	& $f_{2,+}$ 	& $f_{2,+}$ 	& 5076.435	& 0.99	& 196.99 	& 		&		&1 				& 3 		& +1	  	\\
$f_{2,0}$ 	& $f_{2,0}$ 	& $f_{2,0}$ 	& 5073.235	& 2.87	& 197.11 	& 197.60	& 197.62	&1\tablenotemark{*}	& 3 		& 0  	  	\\
$f_{2,-}$ 	& $f_{2,-}$  	& $f_{2,-}$ 	& 5070.03   	& 2.84	& 197.24 	& 		&		&1				& 3 		& $-1$	\\
		& $f_{7}$		& $f_{7}$		& 4398.369	& 0.80	& 227.36	& 226.94	& 227.09	&2				& 8		& 0		\\
$f_{1,+}$ 	& $f_{1,+}$	& $f_{1,+}$	& 4313.369	& 3.61	& 231.84	& 		&		&1				& 4		& +1		\\
$f_{1,0}$ 	& $f_{1,0}$	& $f_{1,0}$	& 4309.919	& 3.15	& 232.02   & 231.88	& 232.16  	&1\tablenotemark{*}	& 4		& 0		\\
$f_{1,-}$ 	& $f_{1,-}$	& $f_{1,-}$	& 4306.520	& 4.29	& 232.21   & 	   	&		&1				& 4		& $-1$	\\
$f_{3,+}$ 	& $f_{3,+}$	& $f_{3,+}$	& 3684.966	& 1.07	& 271.37	&		&		&1				& 5		& +1		\\
$f_{3,0}$ 	& $f_{3,0}$	& $f_{3,0}$	& 3681.800	& 1.61	& 271.61   & 271.52	& 270.71	&1\tablenotemark{*}	& 5  		& 0		\\
$f_{3,-}$ 	& $f_{3,-}$	& $f_{3,-}$	& 3677.996	& 0.54	& 271.89   & 		&	   	&1				& 5  		& $-1$	\\
$f_{4}$ 	& $f_{4}$		& $f_{4,0}$	& 3294.381	& 0.67	& 303.55   & 303.58  	& 304.33	&1				& 6  		& +1		\\
		& 			& $f_{4,-}$	& 3290.0		& 0.30	& 303.95   & 		& 	  	&1\tablenotemark{*}	& 6   	& 0		\\
$f_{5}$ 	& $f_{5}$		& $f_{5}$		& 2658.769	& 1.31	& 376.11   & 376.51	& 374.98  	&1				& 8 		& 0		\\
\hline
		&                	& $f_{9}$ 		& 6207.2    	& 0.25	& 161.10 	& 		& 162.25	&2 				& 5 		& 0		\\
		&			& $f_{10}$	& 4450.0		& 0.26	& 224.72	&		& 		&3?				& 		& 0		\\
		&			& $f_{8}$		& 3746.665	& 0.25	& 266.90	& 		& 267.45	&2				& 10		& 0		\\
\tableline
\end{tabular}
\tablenotetext{1}{Analysis based on Q7 data \citep{Oestensen11}}
\tablenotetext{2}{Analysis based on Q10-Q12 data \citep{Oestensen13}}
\tablenotetext{3}{Analysis based on Q10-Q17 data (this work)}
\tablenotetext{*}{$\ell$ identification fixed by observed rotational splitting}
\end{center}
\end{table}

\subsection{Mode Stability}
\label{runningft}

\begin{figure}
  \includegraphics[scale=1.00,viewport=0 0 50 350]{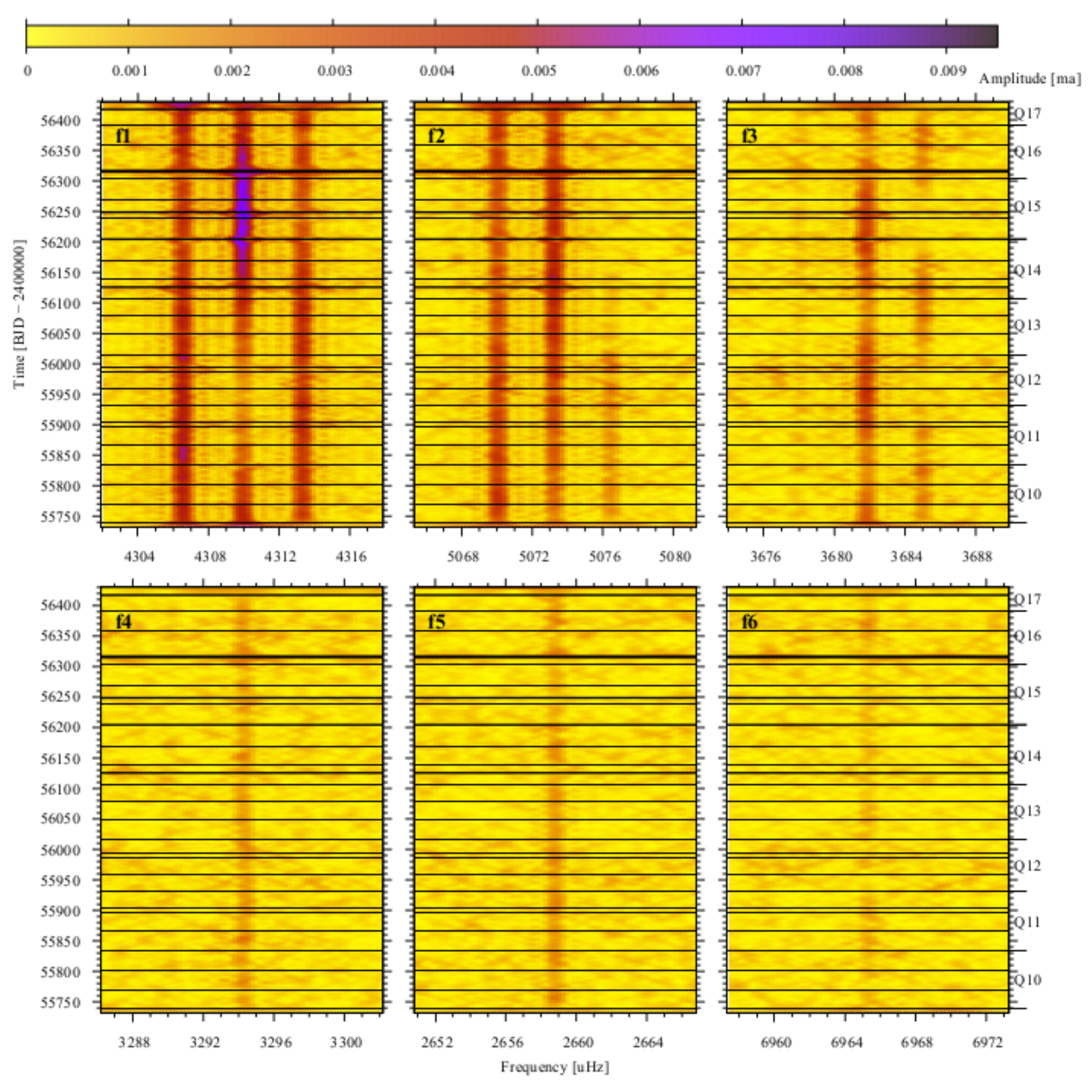}
  \caption{
  \label{f2}
  Sliding Fourier transforms for the six highest-amplitude modes. Time in BJD +2400000 is given on the left-hand axis and \emph{Kepler} mission months are indicated on the right-hand side. Horizontal lines mark the monthly Earth downlink interruptions.
  }
\end{figure}

While we get a number of scientifically interesting results from the asteroseismic study alone, if we want to use the star to measure plasmon neutrino rates, it must have at least one mode that is stable on evolutionary time scales. By ``stable", we mean stable in frequency, not necessarily in amplitude. The cooling rate depends on the frequency alone, not the amplitude of the mode. Sliding Fourier Transforms or sFTs \citep{Jacobsen03} are a graphical way to monitor both the amplitude variations of modes and their frequency stability over time. In Figure \ref{f2} we show the sFTs for the six highest-amplitude modes. The signal-to-noise is too high for the lower amplitude modes to be apparent. 

The sliding FTs were produced by scanning the full Q10-Q17 light curve with a 14-day sliding window, stepped every two days, for an oversampling of seven in time. There is also an oversampling of four in frequency to make the pictures smooth. The peaks of the resulting FTs are lined up, color coded by amplitude, and plotted versus time on the vertical axis. Color or shade variations mark amplitude variations. Modes stable over time appear as straight vertical rails on the figures. The sliding FTs presented in Figure \ref{f2} cover a period of time of two years. Over two years, the modes appear stable in frequency. We discuss more quantitative tests of mode stability in Section \ref{discussion}.

Using the periods identified in Table \ref{t1}, a linear least-squares fit to each month of data shows that two components of the $f_2$ triplet are extremely stable in phase, with an r.m.s. scatter in phase for each very comparable to the mean least-squares phase uncertainties. However, the $f_3$ and $f_4$ modes are far less stable in amplitude and phase, as can be seen by the much broader peaks in the overall FT in Figure \ref{f1}, as well as in the running FT shown in Figure \ref{f2}.

\section{Asteroseismic analysis}
\label{analysis}

\subsection{The Models}
\label{models}

To compute our models, we used the White Dwarf Evolution Code (WDEC). The WDEC evolves hot polytrope models from temperatures above 100{,}000~K down to the temperature of our choice. Models in the temperature range of interest for the present study are thermally relaxed solutions to the stellar structure equations. Each model we compute for our grids is the result of such an evolutionary sequence. The WDEC is described in detail in \citet{Lamb75} and \citet{Wood90}. We used smoother core composition profiles and implemented more complex profiles that result from stellar evolution calculations \citep{Salaris97}. We updated the envelope equation of state tables from those calculated by \citet{Fontaine77} to those given by \citet{Saumon95}. We use OPAL opacities \citep{Iglesias96} and plasmon neutrino rates published by \citet{Itoh96}. 

\begin{figure}
  \includegraphics[scale=0.8]{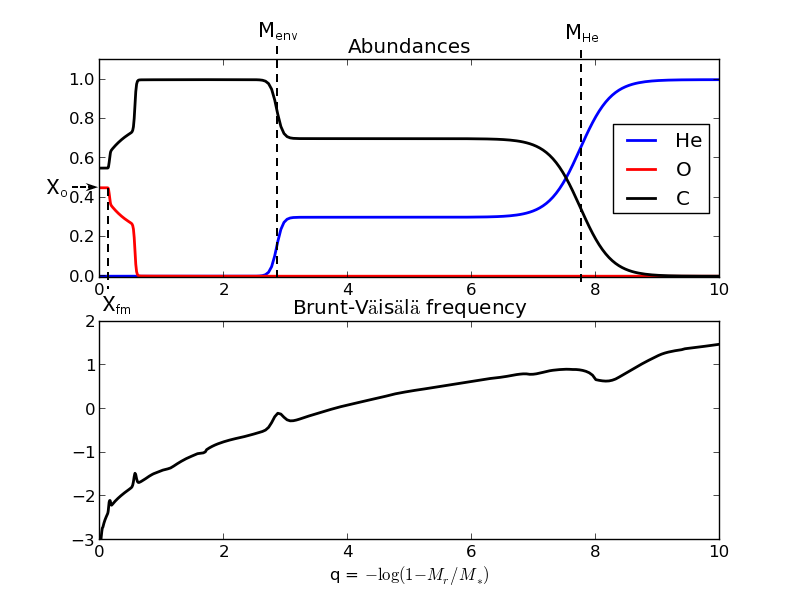}
  \caption{
  \label{f3}
 {\em Upper panel:} Chemical composition profiles of our models to illustrate the parameterization of the internal chemical structure. The center of the model is to the left and the surface to the right. $q=2$ corresponds to $M_r = 0.99 \: M_*$. The vertical axis shows fractional abundances. {\em Lower panel:} The corresponding \bvf frequency ($\log{N^2}$ on the vertical axis). Each chemical transition leads to a bump in the \bvf frequency.
  }
\end{figure}

DBVs are younger than their cooler cousins the DAVs. Time-dependent diffusion calculations \citep[e.g.][]{Dehner95} show that at 24{,}000~K, a typical temperature for a DBV, the carbon has not fully settled into the core of the star yet. We expect the helium layer to be separated into a mixed He/C layer with a pure He layer on top, as shown in Figure \ref{f3}. Following  Metcalfe (2005), we adopted and parameterized this structure in our models. 

We calculated grids of models using the WDEC and then ran a fitting subroutine to match the periods of the models ($P_{\rm calc}$) with the observed periods and calculated residuals using the usual formula

\begin{equation}
\label{eq1}
\sigma_{\rm RMS} = \sqrt{\frac{\sum_{1}^{n_{\rm obs}} {(P_{\rm calc}-P_{\rm obs})^2}}{n_{\rm obs}}},
\end{equation}

\noindent where $n_{\rm obs}$ is the number of periods present in the pulsation spectrum. 

In our models, we only calculated $\ell=1$ and $\ell=2$ modes. $\ell=3$ modes and above suffer from significant geometric cancellation \citep{Dziembowski77} and would have to be driven at very large amplitudes in order to be observed. This is not to say that they cannot be observed. There is photometric evidence for the possible detection of $\ell=4$ modes in G 185 and G 29-38 \citep{Thompson04,Thompson08}. It is also true that in the pulsation spectrum of white dwarfs that have numerous resolved multiplets, we see mostly triplets. A striking example is PG 1159-035 \citep{Costa08}. For that star, 23 of the 29 modes are resolved triplets and the rest are multiplets with the same frequency spacing as the triplets.

Another reason we ignore higher $\ell$ modes is that if we include them, the fits become severly underconstrained. This occurs because the period spacing scales as $1/(\ell\sqrt{\ell+1})$. With a smaller period spacing in the models, it becomes easier to find matching periods. Ignoring higher $\ell$ modes is an assumption we must make in the period fitting but in light of the observations, it is a reasonable assumption to make. For \kicnsp, we are fortunate to have several triplets to guide mode identification.

\subsection{Parameters and grids}
\label{grids}

In our asteroseismic fits, we vary up to six parameters; the effective temperature, the mass and four structure parameters. There are two parameters associated with the shapes of the oxygen (and carbon) core composition profiles: the central oxygen abundance ($X_o$) and the edge of the homogeneous carbon and oxygen core ($q_{\rm fm}$, as a fraction of stellar mass). For envelope structure, $M_{\rm env}$ marks the location of the base of the helium layer and $M_{\rm He}$ the location where the helium abundance rises to $1$ (see Figure \ref{f3}). $M_{\rm env}$ and $M_{\rm He}$ are mass coordinates, defined as e.g. $M_{\rm env} = -\log(1 - M(r)/M_*)$, where $M(r)$ is the mass enclosed in radius r and $M_*$ is the stellar mass. 

In our previous analysis \citep{Bischoff-Kim11b}, we only had five modes and therefore had to make some assumptions and fix some of the parameters. We fixed  $M_{\rm env}$ to -2.80 for all models and $M_{\rm He}$ to -5.50 for models between 20{,}000\ K and 22{,}000\ K, -5.90 for models between 22{,}200~K and 26{,}000~K and -6.30 for models between 26{,}200~K and 28{,}000~K. This was done to match the profiles resulting from the diffusive settling computed by Dehner \& Kawaler (1995).

With at least seven modes, we no longer need to make these assumptions and we allow all six parameters to vary. We started with the grids already computed for our previous analysis \citep{Bischoff-Kim11b} and then computed new grids to find the best fit model. We kept the grid resolution constant, as we determined that it was the correct resolution from the previous work on the star. We computed a total of a little over six million models in the area of parameter space listed in Table \ref{t2}. It is important to note that the region of parameter space indicated in Table \ref{t2} was explored with multiple grids, some of which overlapped. It is also true that not 100\% of that region of parameter space was explored. We increased the resolution of the grids in regions where the best fits were located.

\subsection{Pre-fitting mode identification}
\label{preid}

The choice of periods to match is never trivial, as we almost never have fully resolved triplets and quintuplets for all modes. In asteroseismic fits, we need to use the observed $m=0$ modes, as their frequency is not influenced by stellar rotation. When we have a single peak mode or a doublet, we never know which peak is the $m=0$ component of the mode. In the case of \kicnsp, we are fortunate, as we have three clearly resolved triplets. With the resolved triplets, we learn three things: 1) the triplets are very likely $\ell=1$ modes , 2) what the $\ell=1$ rotational frequency splitting is and by scaling, what the $\ell=2$ rotational frequency splitting is as well, and 3) the central peak is the $m=0$ component of the mode, to use in the asteroseismic fits.

Another factor that makes the asteroseismic analysis of \kic better constrained is that the modes all have relatively short periods. As such, the rotational splitting in period space is very small. The rotational splitting in frequency is related to the orbital period of the star, to first order, by the relation
\begin{equation}
\label{rotsplit}
{\rm \Delta} \sigma = m(1-C_{k\ell})\Omega,
\end{equation}

\noindent where $\Omega$ is the rotation frequency of the star and $C_{k\ell}$ a parameter based on the moment of inertia of the mode. For high-$k$ modes (the asymptotic limit), $C_{k\ell} \approx 1/(\ell + 1)$. The asymptotic limit does not work very well in the present case, as we have low-$k$ modes. We defer the determination of the star's rotation rate from the triplet splitting to our result Section (\ref{fits}). For now, we note that the split in frequency is of order 3-4 $\mu$Hz and that this corresponds to a split in period space of a few tenths of seconds for the modes present in the spectrum (see Table \ref{t1}). This means that for this particular object and for our purposes, the $m=0$ identification is not crucial, but it is nice to know nonetheless. The $\ell$ identification from the triplets and rotational splits of doublets is what is most useful.

From the existence of triplets, we are able to determine that $f_1$, $f_2$, and $f_3$ are $\ell=1$ modes, and also can set the central component as the period to include in our fits. $f_4$ appears as a doublet with our detection limit. By eye, one can see a third component that is trying to rise above the noise. Based on this, we identify the 3294.381 $\mu$Hz frequency as the $m=0$ component of the mode. The spacing of the components of $f_4$ is a little large for an $\ell=1$ mode, but not unreasonably so. We allowed the $\ell$ identification of this mode to vary when perfoming the asteroseismic fits.

We allowed the $\ell$ identification of the other modes to vary between $\ell=1$ and $\ell=2$. In our model fits, we do not consider higher-$\ell$ modes, as we expect them to be greatly reduced in amplitude by geometric cancellations \citep{Dziembowski77}. However, notice in Table \ref{t1} that if $f_{10}$ is real, there are three potentially independent modes between 220 and 235 seconds (224.72\ s, 227.36\ s, and 232.02\ s). This is impossible to accommodate with simply $\ell=1$ and $\ell=2$ modes. The rotational split, as we have mentioned, is of the order of a few tenths of second in this region of the pulsation spectrum and so the three periods belong to three separate modes. The period difference between two consecutive $\ell=1$ modes for this star is of order 36 seconds and for $\ell=2$ modes the figure is about 21 seconds (Section \ref{fits}). No amount of mode trapping (in the current models) can make two $\ell=2$ modes 7 s apart. 

These three periods are too far apart to be rotationally split modes, and too close together to be members of the same $\ell$ sequence. If all three are real and our models not completely wrong, one must be a higher-$\ell$ mode, the most likely being $\ell=3$. So now we have to decide which one is the higher-$\ell$ mode that we leave out of our asteroseismic fits. The 232.02 s mode is clearly identified as an $\ell=1$ mode because it is the central member of a triplet with a frequency spacing consistent with an $\ell=1$ splitting. That leaves the 224.72 s mode and the 232.02 s mode.  Because higher-$\ell$ modes suffer from greater geometric cancellation, we expect $f_7$ to be the $\ell=2$ mode and $f_{10}$ the $\ell=3$ mode.

\subsection{Results of the asteroseismic fitting}
\label{fits}

We performed our initial fit with the seven modes listed in the upper part of Table \ref{t1}. These are the modes that are above the 5-$\sigma$ detection limit. We list the best fit parameters based on matching these modes in Table \ref{t2} as Model 1. As a numerical experiment, we also tried including the modes that were below our detection limit. The best-fit model based on the nine-mode spectrum is labeled as Model 2. We find that two of the three modes that are below the 5-$\sigma$ detection limit (e.g. excluding $f_{10}$, see Section \ref{preid}) fit within the $\ell=2$ sequence quite neatly. The quality of the fit for Model 2 goes down a little, but not significantly so. This may be a sign that our five sigma detection limit is too conservative.

We note that for the five-mode fit \citep{Bischoff-Kim11b}, the $\sigma_{\rm RMS}$ was equal to 0.28 s. But this was based on five periods and four free parameters. In Table \ref{t2}, we also give the Bayes Information Criterion (BIC) number, a statistic that normalizes the quality of fits by number of free parameters and number of constraints. For a discussion applied to this parameter study, see \citet{Bischoff-Kim11b}. For the 5-period fit, the BIC number was equal to -0.41. For Model 1 and Model 2, we find BIC values that are more negative, meaning the fits are better. The $\sigma_{\rm RMS}$ went up a little, while the number of constraints relative to the number of parameters was greater and should have led to a higher $\sigma_{\rm RMS}$ if the fits were of equal quality. In short, these are very good fits.

\begin{table}
\begin{center}
\caption{
\label{t2}
Region of parameters space explored, and best fit parameters. $M_{\rm env}$ is the location of the base of the mixed helium and carbon layer, $M_{\rm He}$ that of the base of the pure helium layer, $X_o$ the oxygen abundance in the center and $q_{\rm fm}$ the location of the edge of the homogeneous C/O core. } 
\begin{tabular}{ccccccccc}
\tableline\tableline
		& $T_{\rm eff}$ (K)	& $M_*$ (Solar) 	& $M_{\rm env}$	& $M_{\rm He}$	& $X_o$	& $q_{\rm fm}$ ($M_*$)	& $ \sigma_{\rm RMS}$	& BIC	\\
\tableline
\multicolumn{9}{c}{Parameter space explored (ranges are inclusive)} \\
Minimum 	& 27{,}550			& 0.500			& $-3.20$			& $-10.00$			& 0.10		& 0.10	\\
Maximum 	& 30{,}400			& 0.585			& $-2.20$			& $-6.80$			& 1.00		& 0.44	\\
Step size	& 100			& 0.005			& 0.10			& 0.10			& 0.05		& 0.02	\\
\tableline
\multicolumn{9}{c}{Best fits, this work} \\
Model 1	& 29{,}650			& 0.550			& $-3.10$			& $-7.90$			& 0.55		& 0.22	& 0.362 s 	& $-12.54$	\\
Model 2	& 29{,}350 			& 0.550			& $-3.00$			& $-8.40$			& 0.70		& 0.22	& 0.741 s 	& $-3.92$	\\
\tableline
Five-period fit\tablenotemark{1} & 29200	& 0.570			& $-2.80$			& $-6.30$			& 0.60, 0.65	& 0.36	& 0.28  s 	& $-0.41$	\\
\tableline
\end{tabular}
\tablenotetext{1}{Reproduced from \citet{Bischoff-Kim11b}}
\end{center}
\end{table}

\begin{figure}
  \includegraphics[scale=0.32]{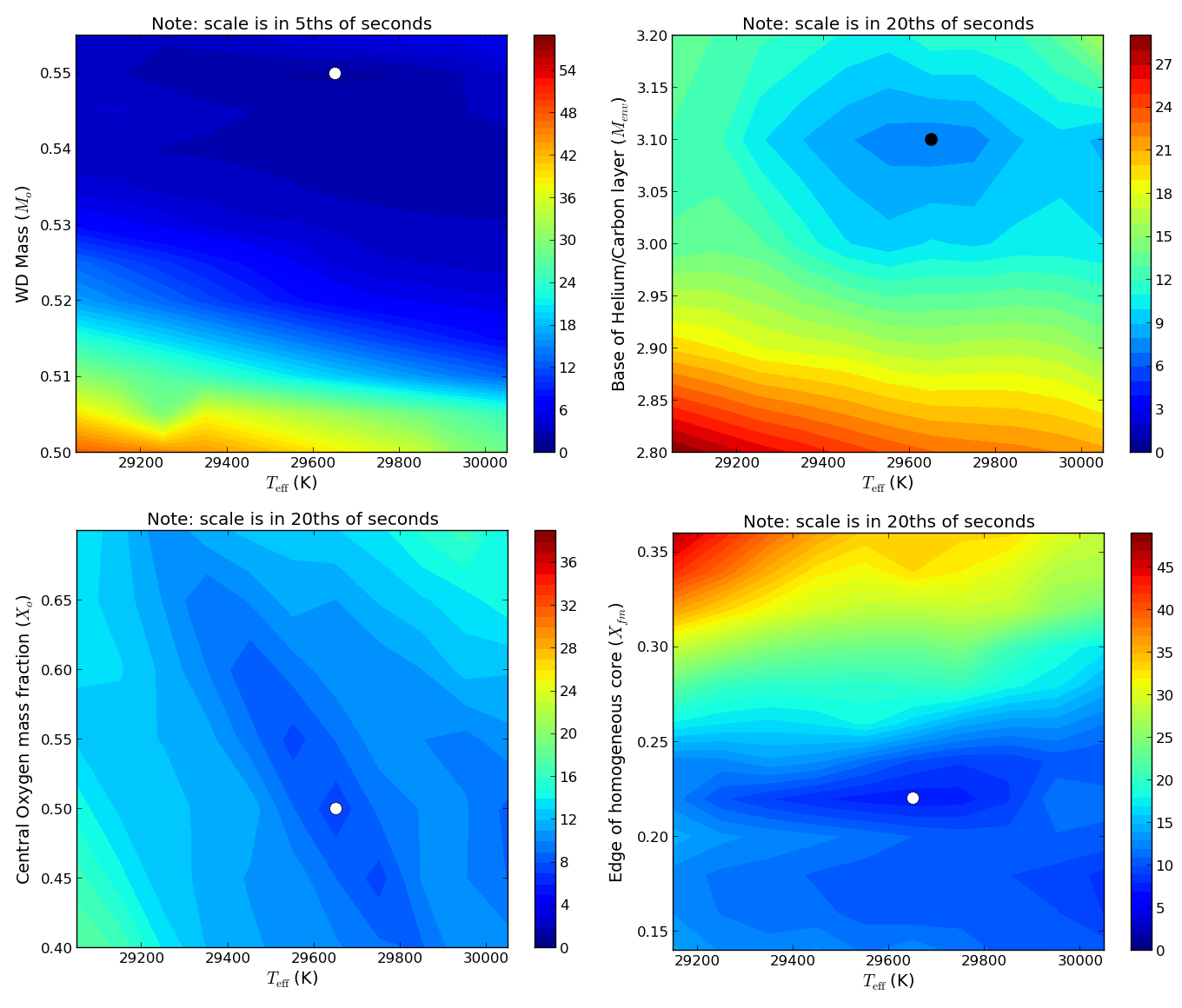}
  \caption{
  \label{f4}
The location of the best fit models in the parameter space explored for four chosen cuts through the 6-dimensional parameter space. The shading corresponds to the $\sigma_{\rm RMS}$ of the models, according to the scale on the right of each plot. In each panel, the dot indicates the location of the best fit model (Model 1 in table \ref{t2}). 
  }
\end{figure}

We show the location of the best fit models in the parameter space explored in Figure \ref{f4}. This is a subset of all the 2D cuts one can make through a 6-dimensional parameter space. The first cut (mass versus effective temperature) can be used to confront the results of the asteroseismic fits against those of spectroscopy, when they become available. The other three plots show three structure parameters: the location of the base of the mixed helium and carbon layer, the central oxygen abundance, and the location of the edge of the homogeneous C/O core. Note the sharply defined region of best fit in the latter. This is the reason a smaller step size is required for this parameter ($q_{\rm fm}$ in table \ref{t2}). 

The period-by-period match and mode identification are listed in Table \ref{t1}. While the mode identification for $f_5$ for Model 1 is clearly $\ell=1$, other models that come not far behind in quality of fit have an $\ell=2$ mode that match that mode as well. $f_5$ happens to fit in both $\ell$ sequences and we indicate both possibilities in Table \ref{t1}. Based on our results, we can recalculate an $\ell=1$ and an $\ell=2$ period spacing for \kicnsp. We have four consecutive $\ell=1$ modes and if we include $f_5$ and the modes that are below the 5-$\sigma$ detection limit, five $\ell=2$ modes. We find an average period spacing of 36.0 s for the $\ell=1$ modes (35.9 s if we include $f_5$ in the sequence) and 21.2 s for the $\ell=2$ modes (regardless of which of $f_7$ or $f_{10}$ we include in the sequence). This refines the 35.7 s and 20.6 s quoted in \citet{Oestensen13}.

With our best-fit model, we can use the rotational splitting of some of the modes to determine a rotation rate for the star. A preliminary analysis based on the asymptotic limit ($C_{k\ell} \approx \ell/(\ell+1)$) suggested a rotation period of 1.7 days \citep{Oestensen13}. A refined analysis based on the models gives us $C_{k\ell}$ values of 0.429 (lowest $k$ mode) to 0.478, as opposed to 0.5 in the asymptotic limit. We include in our analysis modes $f_1$ to $f_4$. Based on these modes and equation \ref{rotsplit}, we find a rotation period of $1.8 \pm 0.4$ days, consistent with the preliminary result. 

\section{Discussion and Conclusions}
\label{discussion}

We confirm and refine the best fit parameters that we found based on the five-period fits of \kicnsp. The internal structure parameters have fluctuated some, as is expected, since we added constraints and allowed more parameters to vary. The pulsations clearly point to the star being near the blue edge of the DBV instability strip. One consistent feature of the models is how thin the pure helium layer is (as evidenced by the small value of $M_{\rm He}$). This is in contrast with two other DBs of similar mass analyzed with the same models, CBS 114 \citep{Metcalfe05b} and EC 20058-5234 \citep{Bischoff-Kim11a}. We continue to see evidence that the hotter the DB model, the thinner the pure helium layer, supporting the theory that as a DB cools and helium diffuses outward, the pure helium layer becomes thicker. For CBS 114 ($T_{\rm eff}=24{,}900$~K), $M_{\rm He}=-5.94$, while for EC20058-5234 ($T_{\rm eff}=29{,}200$~K), $M_{\rm He}=-6.10$. For \kicnsp, the pure helium layer is thinner than both, $M_{\rm He}< -8.0$. \kic is also hotter than both  ($T_{\rm eff}=29{,}550$~K), confirming the trend.

The quality of the fit for \kic is so far the best we have managed with our methods. This is most likely because of the rich multiplet structure we have for \kic, thanks to the exceptional \emph{Kepler} data. We have very little a priori mode identification information for the other DBVs we have studied. Finding a global minimum in asteroseismic fitting is always a challenge. We may be able to use what we have learned from the asteroseismic analysis of \kic to better constrain EC20058-5234, since the two stars are close pulsational cousins. For instance, EC20058-5234 has a 195.0 s mode that we identified from our fits as an $\ell=2$ mode. In \kicnsp, there is a 197.11 s mode that is the central peak of an $\ell=1$ triplet. There are other similarities in the two pulsation spectra we could exploit.

More quantitative tests are required in order to determine the potential for each mode to measure a cooling rate of the object. The size of the signal we are looking for is $\frac{{\rm d}P}{{\rm d}t}\sim 10^{-13}  \;{\rm s\cdot s^{-1}}$ \citep{Corsico04}. This translates to a change in a given period of one second in 300,000 years. A method based on counting cycles (the ``O-C" method) allows us to perform such a measurement in a reasonable amount of time, less than 6 years. The rate of cooling of the DAV G117-B15A ($\approx 10^{-15} \;{\rm s\cdot s^{-1}}$), the first measurement of its kind, was made based on a 30-year baseline of pulsation data \citep{Kepler05a}. For \kicnsp, it will take less time because the rate of period change is 100 times greater. Preliminary analyses by the O-C method of the three triplets ($f_1$, $f_2$ and $f_3$) show that at least two components of $f_2$ are stable on evolutionary time scales.

By securing two additional modes for a total of seven independent pulsation modes, we are able to conservatively perform a six-parameter asteroseismic study of \kic and find a best-fit model that matches the period spectrum with unprecedented accuracy. We confirm previous studies that showed that \kic is hotter than its close pulsational cousin EC 20058-5234. We find a thin pure helium layer, consistent with a star not as far along in its evolution and the outward diffusion of helium. We eagerly await the results of spectroscopic studies, both in the optical and ultraviolet, as well as those from light curve fitting to verify our results. We also confirm a rotation rate of $1.8 \pm 0.4$ days.

Another noteworthy result is the possible discovery of an $\ell=3$ mode in the pulsation spectrum. So far, it has always been possible to fit entire pulsation spectra of white dwarfs with only $\ell=1$ and $\ell=2$ modes. That is not to say that there are no higher $\ell$ modes hidden among those, but we generally consider the detection of $\ell=3$ modes unlikely, as we expect them to have very low optical pulsation amplitudes. 

In the case of \kicnsp, if $f_{10}$ is real, then we cannot explain the existence of three modes separated by a few seconds without invoking a higher-$\ell$ mode. The modes are too close to be rotationally split modes, and too far to be consecutive modes in the same $\ell$ sequence. Two other modes with amplitudes slightly lower than that of $f_{10}$ fit well in the $\ell=2$ sequence so even though $f_{10}$ falls below the chosen 5-$\sigma$ detection  limit, it may still be real. $f_{10}$ is not a combination of any of the other observed modes.

Space-based observations have allowed us to detect a rich pulsation spectrum in \kic that would not have been possible from the ground. Thanks to {\em Kepler}, we have obtained enough information to perform a well-constrained asteroseismic study of this hot DBV. We can continue to monitor this object from the ground to monitor the period change of the pulsations. This will give us an unprecedented measurement of its cooling rate, which will finally allow us to empirically evaluate the contribution plasmon neutrinos have on white dwarf cooling rates.

\acknowledgments

The authors gratefully acknowledge the {\em Kepler} team and everyone who made the mission possible. Funding for the {\em Kepler} mission is provided by the NASA Science Mission Directorate. J.J.H. acknowledges funding from the European Research Council under the European Union's Seventh Framework Programme (FP/2007-2013) / ERC Grant Agreement n. 320964 (WDTracer). J.L.P. acknowledges the support of the Crystal Trust and Mt. Cuba Observatory. The source code for Figure \ref{f4} was kindly provided by Fergal Mullally.

\bibliographystyle{apj}
\bibliography{index}


\end{document}